\documentclass[12pt]{iopart}
\usepackage{graphicx}
\usepackage{latexsym}
\usepackage{amsthm}

\usepackage{iopams}
\usepackage{url}
\def\aap{A\&A\,  }
\def\apjs{ApJS  }
\def\apss{Astrophysics and Space Science  }
\def\araa{ARA\&A  }

\def\mnras{MNRAS\,  }

\def\prd{Phys. Rev. D   }
 
\def\ssr{Space Science Reviews} 
\def\snr{SN\,1993J~}
\def\grb{GRB 161214B~}
\def\secondogrb{GRB 050814~}
\def\h0units{\mathrm{km\,s^{-1}\,Mpc^{-1}}}

\newcommand{\om}{\Omega_{\rm M}}

\newcommand{\ola}{\Omega_{\Lambda}}

\begin{document}
\title
{
Classical and  relativistic models for  time duration 
of  gamma-ray bursts
}
\vskip  1cm
\author     {L. Zaninetti}
\address    {
Physics Department,
 via P.Giuria 1,\\ I-10125 Turin,Italy 
}
\ead {zaninetti@ph.unito.it}

\begin {abstract}
A classical model based on a power law assumption for 
the radius-time relationship in  the expansion of a
Supernova  (SN)  allows to derive an
analytical expression for the flow of mechanical kinetic energy
and the time duration of Gamma-ray burst (GRB).
A random process based on the ratio  of two truncated 
lognormal distributions for luminosity and luminosity distance 
allows to derive  the statistical distribution 
for time duration of GRBs.
The high velocities involved in the first phase 
of expansion of a SN requires a relativistic 
treatment.
The circumstellar medium is
assumed to follow  a density profile of
Plummer type with eta=6.
A series solution for  the relativistic  flow of kinetic 
energy allows to derive in a numerical way 
the duration time for GRBs.
Here we analyse  two cosmologies: the standard cosmology 
and the plasma cosmology.
\end{abstract}
\vspace{2pc}
\maketitle 


\section{Introduction}
The theoretical efforts   for gamma-ray bursts (GRBs)
analyze: (i) the different predictions between cannonball
and fireball model,  see \cite{Dado2016,Willingale2017},
(ii) the acceleration of 
ultrahigh-energy cosmic rays (UHECRs) 
at  the afterglow phase, see \cite{Asano2016},
(iii) a possible connection with 
High-energy neutrinos (HEN) and gravitational waves (GW),
see \cite{Moharana2016},
(iv) the frames of Hypersphere World-Universe
Model (WUM), see \cite{Netchitailo_2017}.
We briefly recall that 
the time duration for Gamma-ray burst (GRB) 
is  measured without the knowledge  
of the redshift, see as an example \cite{Bhat2016}.
The distance of GRBs is therefore unknown and the
galactic or extra-galactic origin should be analyzed.
We test therefore in the following 
the reliability of time duration for GRBs along 
two astrophysical hypothesis: 
(i) the GRBs are generated 
in external galaxies,
(ii) the GRBs are connected  with
the light curve (LC)   of  a Supernova  (SN).
A first classification of time duration  
provides a division between  short and long GRBs,
with the boundary at $\approx$ 2\,s, see \cite{Berger2014}.
The form  of the  probability density function 
(PDF)
which produces the better fit to the sequence 
of times is also subject of research,
as an example  a two  Gaussian-fits 
has been suggested  by \cite{Tarnopolski2015}  
and a lognormal PDF by \cite{Horvath2016}.

This paper reviews in Section \ref{secpreliminaries}   
the current status of the observations 
for the time duration of GRBs.
Section  \ref{secsimple} develops a simple classic 
model for theoretical time duration for GRBs.
Section  \ref{secsimulation}  contains a simulation
for time duration of GRBs  
based on the random generation of  luminosity
and distance luminosity.
Section \ref{secrelativistic} 
derives a relativistic equation of 
motion for SN, 
deduces the relativistic flow of energy
and finally evaluates the time duration for GRBS in a relativistic
environment.

\section{Preliminaries}

\label{secpreliminaries}
This section reviews the observations   
of the Fermi Gamma-ray Burst Monitor (GBM) for the time duration 
and fits the data with the truncated lognormal PDF.

\subsection{The observations}

\begin{figure}
\begin{center}
\includegraphics[width=10cm]{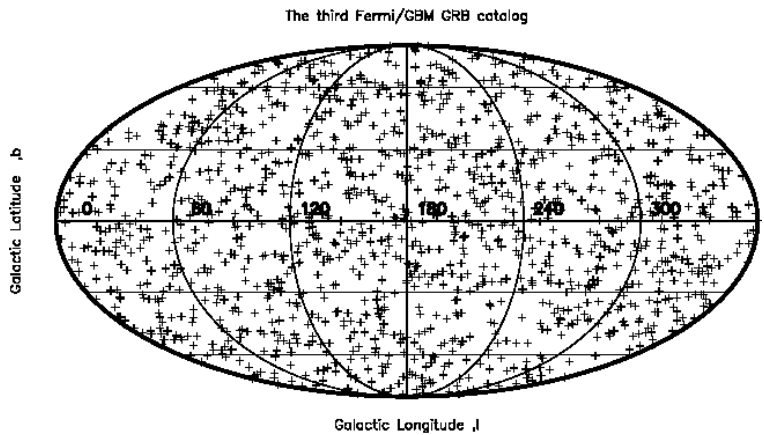}
\end{center}
\caption
{
Sky distribution of  GRBs
in Galactic coordinates projected using the Mollweide projection.
}
\label{fermi_projection}
\end{figure}

Most of the typical GRBs have two stages: the initial prompt 
emission phase
observed in the gamma-ray band and the following afterglow phase with
multi-bands' emission. 
The observational time duration is measured during
the prompt emission phase with the typical energy band between tens of keV
to hundreds of keV.  Observationally, the light curve of the afterglow phase
can be fitted by a broken power-law, but the light curve during the prompt  
emission phase is highly variable and completely different from the
afterglow phase.
The time duration  of GRBs 
has been monitored by GBM in the
$50-300$  keV energy range,
see the third catalog \cite{Bhat2016} 
which  is available on line at \url{http://cdsweb.u-strasbg.fr/}.
The above catalog reports two times of burst duration 
$T_{50}$   
and $T_{90}$  
which are defined 
as  the interval between the times where the burst
has reached $25\%$ and  $75\%$  of its maximum fluence.
Figure \ref{fermi_projection} 
reports  the sky distribution of  GRBs
in Galactic coordinates.

A simple test on the isotropy of the arrival direction 
of GRBs can be performed dividing the 
the surface area of a sphere of unit radius in $N \times N$ boxes
of equal area.
We then choose $N$ in order to have $\approx\,7 $ theoretical events,
$n_{th}$,  
assuming a uniform distribution on the total surface area,
which means $N=14$.
We now evaluate  the observed averaged number of  GRBs in each box, $n$, 
which turns out to be $n=7.16\pm 2.7$.
The    goodness of the isotropy 
is evaluated
through the percentage error, $\delta$,
which is
\begin{equation}
\delta = \frac{\big | n_{th} - n  \big |} 
{n_{th} } \times 100  \quad =2.4 \%  
\quad .
\end{equation}
The low  value of the percentage error for  the isotropy
allows to state that:
\begin{itemize}
\item the GRBs have an extra-galactic origin
\item the spatial distribution of GRBs 
      is isotropic.
\end{itemize} 

\begin{figure}
\begin{center}
\includegraphics[width=10cm]{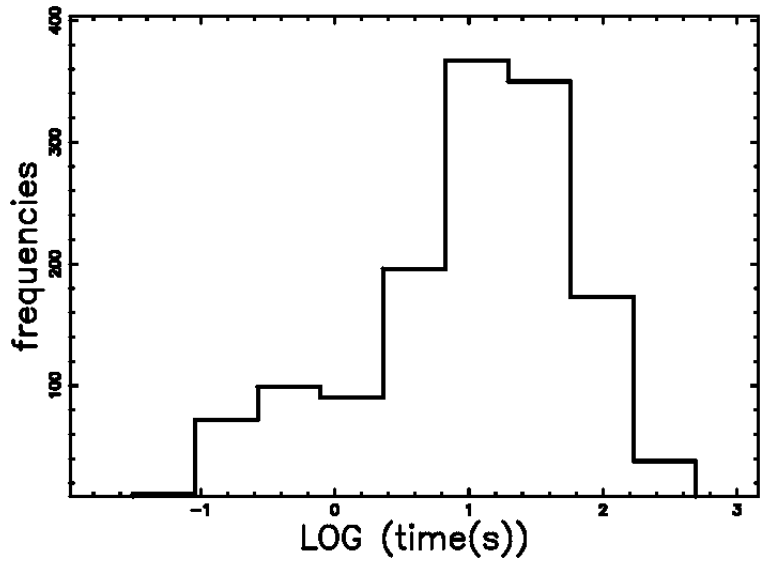}
\end{center}
\caption
{
Frequencies distribution (linear scale)) of 
$T_{50}$  (decimal logarithm scale) for GRBs.
}
\label{histotimet50}
\end{figure}

\begin{figure}
\begin{center}
\includegraphics[width=10cm]{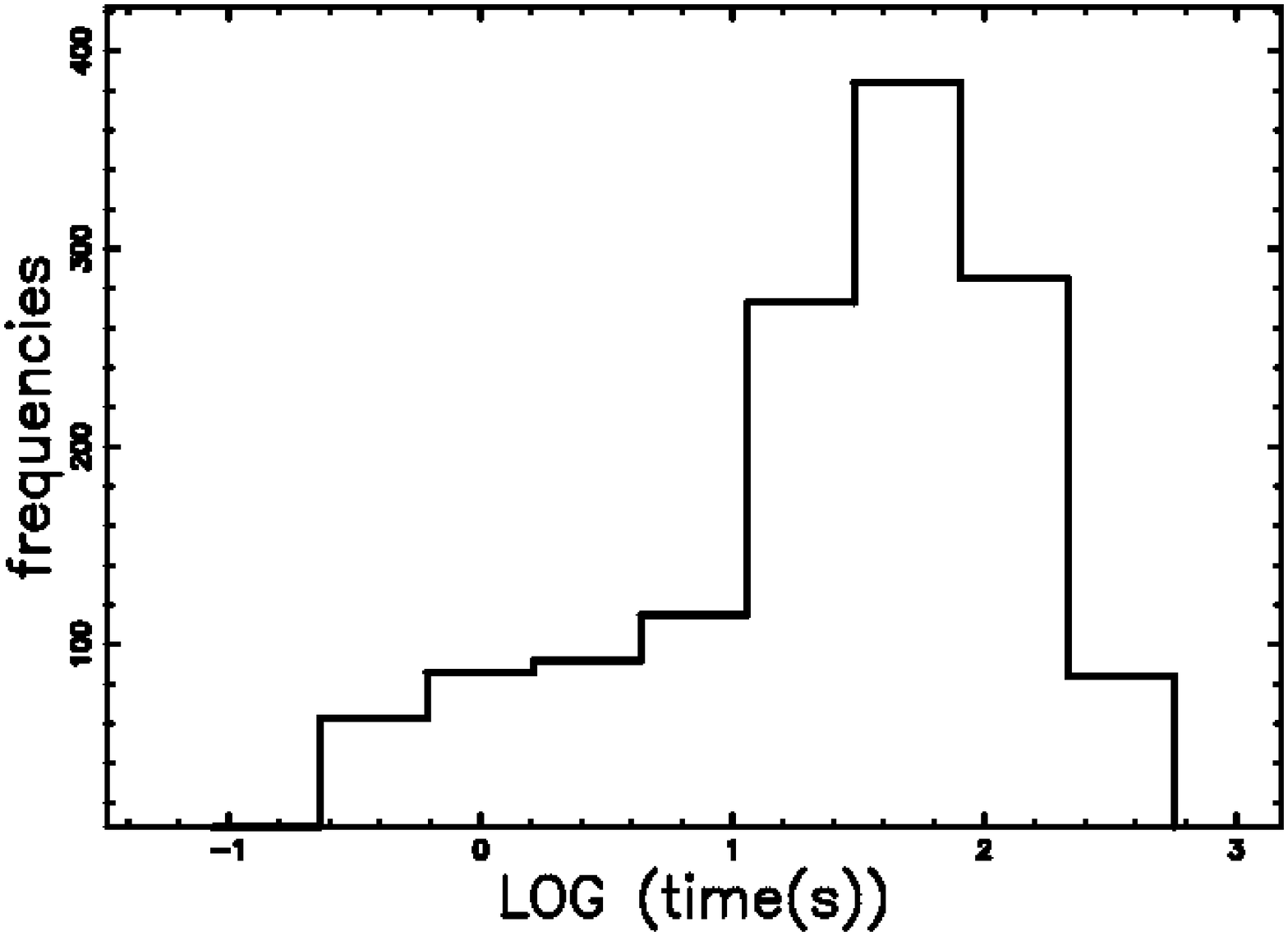}
\end{center}
\caption
{
Frequencies distribution 
(linear scale)) of $T_{90}$  
(decimal logarithm scale) for GRBs.
}
\label{histotimet90}
\end{figure}

The histogram of time distribution for GRBs
for  the case of $T_{50}$ 
in Figure \ref{histotimet50}
and 
in Figure \ref{histotimet90}
for  the case of $T_{90}$ 

Tables \ref {table_time50}   and \ref {table_time90}
report  the main statistical parameters of 
$T_{50}$ and $T_{90}$ respectively.
\begin{table}
 \caption[]
{
The sample parameters    
of $T_{50}$  for GRBs in seconds.
}
 \label{table_time50}
 \[
 \begin{array}{ll}
 \hline
Parameter       & value~(seconds)   \\ \noalign{\smallskip}
 \hline
 \noalign{\smallskip}
elements        & 1405    \\
maximum         & 736.51  \\
mean            & 18.23   \\
minimum         & 0.016   \\
standard ~deviation        & 41.22 \\  
 \hline
 \end{array}
 \]
 \end {table}

\begin{table}
 \caption[]
{
The sample parameters    
of $T_{90}$  for GRBs in seconds.
}
 \label{table_time90}
 \[
 \begin{array}{ll}
 \hline
Parameter       & value~(seconds)  \\ \noalign{\smallskip}
 \hline
 \noalign{\smallskip}
elements        & 1405    \\
maximum         & 828.67  \\
mean            & 39.64  \\
minimum         & 0.048   \\
standard ~deviation        & 62.55 \\  
 \hline
 \end{array}
 \]
 \end {table}

\subsection{The fit}

The high number of decades, $\approx$ 4,
covered by the distribution of $T_{50}$ and   $T_{90}$
suggests to fit the data with the 
truncated lognormal  (TL) PDF
\begin {eqnarray}
TL(x;m,\sigma,x_l,x_u) = \nonumber  \\
\frac
{
\sqrt {2}{{\rm e}^{-\frac{1}{2}\,{\frac {1}{{\sigma}^{2}} \left( \ln  \left( {
\frac {x}{m}} \right)  \right) ^{2}}}}
}
{
\sqrt {\pi}\sigma\, \left( -{\rm erf} \left(\frac{1}{2}\,{\frac {\sqrt {2}}{
\sigma}\ln  \left( {\frac {x_{{l}}}{m}} \right) }\right)+{\rm erf}
\left(\frac{1}{2}\,{\frac {\sqrt {2}}{\sigma}\ln  \left( {\frac {x_{{u}}}{m}}
 \right) }\right) \right) x
}
\label{lognormaltruncated}
\quad ,
\end {eqnarray}
where 
$x$ is the random variable,
$x_l$ is the lower  bound,
$x_u$ is the upper  bound,
$m$   is the  scale parameter
and 
$\sigma$ is the shape parameter,
see \cite{Zaninetti2016c} for more details.
Tables  
\ref{table_time50_tl} 
and
\ref{table_time90_tl}
report the four parameters of  the TL
for $T_{50}$ and $T_{90}$ respectively.
Table \ref{table_time90_tl} reports also 
the parameters of the lognormal PDF 
as well 
the maximum  distance, $d_{max}$, between the theoretical
and the observed   
distribution
function (DF)
in the 
Kolmogorov--Smirnov test (K--S),
see \cite{Kolmogoroff1941,Smirnov1948,Massey1951}.

The DF  is
\begin{equation}
DF(x;m,\sigma,x_l,x_u)=
\frac
{
-{\rm erf} \left(\frac{1}{2}\,{\frac {\sqrt {2}}{\sigma}\ln  \left( {\frac {x
}{m}} \right) }\right)+{\rm erf} \left(\frac{1}{2}\,{\frac {\sqrt {2}}{\sigma}
\ln  \left( {\frac {x_{{l}}}{m}} \right) }\right)
}
{
{\rm erf} \left(\frac{1}{2}\,{\frac {\sqrt {2}}{\sigma}\ln  \left( {\frac {x_{
{l}}}{m}} \right) }\right)-{\rm erf} \left(\frac{1}{2}\,{\frac {\sqrt {2}}{
\sigma}\ln  \left( {\frac {x_{{u}}}{m}} \right) }\right)
}
\quad .
\label{dftruncatedlognormal}
\end{equation}
A random generation of the variate X can be  found
by solving  the following 
non linear equation 
\begin{equation}
R = DF(X;m,\sigma,x_l,x_u) 
\quad ,
\end{equation}
where $R$ is the unit rectangular variate.
\begin{table}
 \caption[]
{
The      TL  parameters    
of   $T_{50}$  for GRBs.
}
 \label{table_time50_tl}
 \[
 \begin{array}{ll}
 \hline
Parameter       & value   \\ \noalign{\smallskip}
 \hline
 \noalign{\smallskip}
elements        & 1405    \\
x_l             & 0.016   \\
x_u             & 736.51  \\
m               & 5.33    \\
\sigma          & 1.89    \\
 \hline
 \end{array}
 \]
 \end {table}

\begin{table}
 \caption[]
{
The TL PDF and lognormal PDF  
parameters  of  $T_{90}$  for GRB.
}
 \label{table_time90_tl}
 \[
 \begin{array}{lll}
 \hline
Parameter       & value~TL & value~lognormal   \\ \noalign{\smallskip}
 \hline
 \noalign{\smallskip}
elements        & 1405    & 1405   \\
x_l             & 0.048   & 0      \\
x_u             & 828.6   & \infty \\
m               & 14.62   & 13.57  \\
\sigma          & 1.911   & 1.8    \\
d_{max}         & 0.096   & 0.1    \\  
 \hline
 \end{array}
 \]
 \end {table}

\section{A classical model for the light curve}

\label{secsimple}
We assume that  the observed
radius--time  relationship, $R(t)$,   
for SN has a   power law
dependence  of the type
\begin{equation}
R(t) = C \times  t^{\alpha}
\label{rpower}
\quad ,
\end{equation}
where $C$ is a constant and $\alpha$
a parameter which  can be fixed by the
observation of the temporal evolution of the radius of 
a SN, in the case of \snr\,   
$\alpha=0.82$,  see \cite{Zaninetti2011a}. 
At time $t_0$ the radius $R_0$ is 
\begin{equation}
R_0(t) = C \times  t_0^{\alpha}
\label{rpower0} 
\quad .
\end{equation}
The velocity, $V(t)$, is
\begin{equation}
V(t) = C\times  \alpha t^{(\alpha-1)}
\quad ,
\label {vpower}
\end{equation}
and  the velocity at time $t_0$ is
\begin{equation}
V_0(t) = C \times  \alpha t_0^{(\alpha-1)}
\quad .
\label {vpower0}
\end{equation}
In classical physics the  density of kinetic energy, $K$,
is
\begin{equation}
K = \frac{1}{2}\rho   V^2
\quad,
\end{equation}
where $\rho$ is the density and $V$ the velocity.
In presence of an area $A$ and when the velocity 
is  perpendicular
to that area,
the mechanical flux of kinetic energy
$L_m$ is
\begin{equation}
L_m = \frac{1}{2}\rho A  V^3
\quad,
\end{equation}
which in SI  is measured in $W$  and in CGS in erg $s^{-1}$
see formula (A28) in
\cite{deyoung}. 
In our  case, $A=4\pi R^2$, which means
\begin{equation}
L_m = \frac{1}{2}\rho 4\pi R^2 V^3
\quad .
\end{equation}
The density in the advancing  layer as function of the radius, 
$\rho(R)$, 
is  assumed
to scale as  
\begin{equation}
\rho(R) = \rho_0 (\frac{R}{R_0})^{-d}
\quad ,
\label{rhor}
\end{equation}
where $\rho_0$ is  the density at radius $R_0$.
According to previous formula the scaling law 
for the mechanical luminosity 
as function of the time
is 
\begin{equation}
L_m = L_{m,0} 
\bigl( \frac{t}{t0} \bigr )^{5\,\alpha -d \alpha -3}
\quad ,
\end{equation}
where $L_{m,0}$ is the mechanical luminosity
at $t=t_0$. 
The observed luminosity at a given frequency $\nu$, $L$,
is  assumed 
to be proportional to the mechanical luminosity
\begin{equation}
L  = \epsilon L_m \quad ,
\label{eqnl}
\end{equation}
where  $\epsilon$ is a constant of conversion.
The luminosity at a   given range of energy   
is expressed in $\frac{erg}{s}$ in CGS 
and $W$ in SI.
As a useful example  
the astrophysical version 
of the luminosity at $t=t_0$  is
\begin{equation}
L= 2.68521\,{\it n_0}\,
({{\it R_{0,pc}}})^{2}({\frac{V_0}{c}})^3 \, 10^{32}
\, W
\quad ,
\label{lsimpleastro}
\end{equation}
where 
$R_{0,pc}$ is the radius    at $t=t_0$ in pc,
$V_0$      is the velocity  at $t=t_0$ in $\frac{km}{s}$,
$c$        is the light  velocity in $\frac{km}{s}$ 
and
$n_0$ is  the number density expressed  in 
particles~$\mathrm{cm}^{-3}$
(density~$\rho_0=n_0m$, where $m=m_{\mathrm {H}}$).  

\begin{figure}
\begin{center}
\includegraphics[width=10cm]{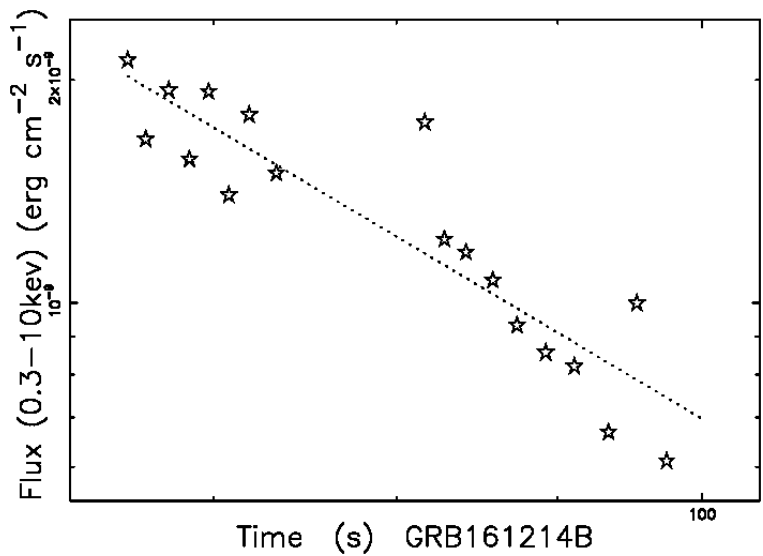}
\end{center}
\caption
{
Light curve of 
\grb  in the first 100 seconds 
(empty stars)
and power law fit (dotted line).
}
\label{grb161214B}
\end{figure}

The observed flux, $F$,   as  a function of the luminosity 
distance,  $D_L$,
is
\begin{equation}
F = \frac{L}{4 \pi D^2}
\quad ,
\label{fluxobserved}
\end{equation}
or
\begin{equation}
F(L,D_L)= \frac
         {
       L_0 \bigl( \frac{t}{t0} \bigr )^{5\,\alpha -d \alpha -3}
         }
   {4 \pi D_L^2}
\quad , 
\label{fld}
\end{equation}
where  $L_0=\epsilon L_{m,0}$.
The observed  flux at a   given range of energy   
is expressed in $\frac{erg}{s\,cm^2}$ in CGS 
and $\frac{W}{m^2}$ in SI.
An on line  collection of   light curves for GRBs is  
made by the Swift GRB  Mission and 
can be found  at 
\url{http://swift.gsfc.nasa.gov/index.html},
see also \cite{Evans2007,Evans2009}:
the band of the observations is (0.3-10) kev.
As a practical example we processed  
\grb, see \cite{Krimm2016},
with review  data in Table \ref{table_GRB161214B},
and real  data available at
\url{http://www.swift.ac.uk/xrt_curves/00726885/}.
Figure \ref{grb161214B} displays  
the LC of such burst  as well a power law fit.
\begin{table}
 \caption[]
{
Some parameters    
of \grb.
}
 \label{table_GRB161214B}
 \[
 \begin{array}{ll}
 \hline
Parameter       & value   \\ \noalign{\smallskip}
 \hline
 \noalign{\smallskip}
date            & 16/12/14             \\
T_{50} (s)      & 9.67                 \\
T_{90} (s)      &  24.82               \\
C_k ~ t < 100\, s         &  8.43\times 10^{-5}  \\ 
k   ~~~ t < 100\, s         & -2.54               \\            
 \hline
 \end{array}
 \]
 \end {table}
A  first numerical analysis of the observed  
flux, $F_0$,   versus   time relationship 
can be done by assuming  a  power law 
dependence  
\begin{equation}
F_0(t) = C_k t^{k}  \,erg\, s^{-1} 
\label{ratetime}
\quad ,
\end{equation}
where the two parameters $C$ and $k$  are 
found from a numerical analysis of the data,
see Table \ref{table_GRB161214B}.
We now match the observed value of 
$k$ with the theoretical 
exponent given  by equation (\ref{fld})
\begin{equation}
k = 5\,\alpha -d \alpha -3
\quad .
\end{equation}
The above equation allows to deduce 
$d$ once $\alpha$ and $k$ are given: 
as an example,
when $\alpha=0.82$ and $k=-2.54$, 
$d=4.44$. 
The decreasing flux will be visible until a minimum value 
of threshold in the flux 
$F_T$  is reached.
As an example  the instrument Windowed Timing (WT)
of  X-ray
telescope (XRT) on the Swift satellite
has a threshold value of 
$F_T \approx 3.1 \frac{fW}{m^2}$. 
Therefore the GRB will be visible 
up to a maximum value in time of $t_{max}$ 
\begin{equation}
t_{max} = t_0 (\frac{F_T 4 \pi D_L^2}{L_0} 
)^{\frac{1}{5\,\alpha -d \alpha -3}}
\quad ,  
\label{timegrb}
\end{equation}
or 
\begin{equation}
t_{max} = t_0 (\frac{F_T}{F_0} 
)^{\frac{1}{5\,\alpha -d \alpha -3}}
\quad ,  
\end{equation}
where $F_0$ is the observed flux at $t=t_0$.

\section{The simulation}

\label{secsimulation}

A simulation of the duration time for GRBs 
according to equation \ref{timegrb}
requires 
the ratio of two two lognormal PDFs 
( one for       $L_0$ and the 
other one for   $D_L$) 
and 
a cosmological environment.

\subsection{The ratio of two lognormal PDFs}

\label{ratiolognormals}

We now evaluate  the mean and the variance of the ratio 
of two lognormal distributions , $\frac{X}{Y}$.
From the fact that $\ln(\frac{X}{Y})=\ln(X)-\ln(Y)$ and 
 $X$ and $Y$ are lognormally distributed,
it turns out that $\ln(X)$   and $\ln(Y)$ 
are  distributed as a normal PDF.
We assume  that  $\ln(X)$ 
and $\ln(Y)$  have means 
$\mu_X$ and 
$\mu_Y$, 
variances $\sigma^2_X$ and
          $\sigma^2_Y$
, and covariance $\sigma^2_{XY}$
(equal to zero if X and Y are independent) 
and are jointly normally
distributed. 
The difference $Z$ is then  distributed as a normal PDF with mean
$\mu_Z=\mu_X-\mu_Y$ 
and variance 
$\sigma^2_Z$ = $\sigma^2_X$ + $\sigma^2_Y$ - 2 $\sigma^2_{XY}$.

Note that $\frac{X}{Y}=\exp (Z)$ , which means
that $\frac{X}{Y}$ is distributed as a lognormal PDF
with parameters $\mu_Z$ and $\sigma^2_Z$.

\subsection{Cosmological models}

We deal with two cosmologies: 
the $\Lambda$CDM cosmology
and the plasma cosmology.
The the $\Lambda$CDM cosmology is characterized  
by  three  parameters which are the Hubble constant, $H_0$, 
expressed in  $\h0units$ and 
the two  numbers $\om$ and $\ola$,
see Table \ref{cosmologyvalues}.
\begin{table}[ht!]
\caption
{
Numerical values for parameters of the  two cosmologies.
}
\label{cosmologyvalues}
 \[
\begin{array}{ccccc}
\hline
cosmology & compilation   &  H_0~in~\h0units  &  \om     & \ola \\
\hline
\Lambda\!CDM  & ~ Union~2.1 & 69.81 & 0.239 & 0.651        \\
plasma       & ~ Union~2.1 & 74.2  &  ~ & ~        \\
\hline
\end{array}
\]
\end{table}

A simple expression 
for the luminosity distance, $D_l$, in 
$\Lambda$CDM
is obtained  with the 
minimax approximation ($p=3,q=2$) 
\begin{eqnarray}
D_{\rm L,3,2}= \nonumber  \\
\frac
{
0.3597252600+ 5.612031882\,z+ 5.627811123\,{z}^{2}+ 0.05479466285\,{z
}^{3}
}
{
0.0105878216+ 0.1375418627\,z+ 0.1159043801\,{z}^{2}
}
.
\end{eqnarray}
More details on 
the analytical derivation of the luminosity distance
in terms  of  Pad\'e  approximant can be found 
in  \cite{Zaninetti2016a}. 

The plasma cosmology is characterized by one 
parameter which is  $H_0$
and by a simple formula for the luminosity distance 
which is the same of the distance, $d_p$,  
\begin{eqnarray}
D_L(z;H_0) \equiv  d_p(z;H_0)=  \nonumber \\
\frac 
{
 0.359725+ 5.61203\,z+ 5.62781\,{z}^{2}+ 0.0547946\,{z
}^{3}
}
{
 0.010587+ 0.137541\,z+ 0.115904\,{z}^{2}
} 
\quad ,
\label{distancenltired}
\end{eqnarray}
 see
\cite{Brynjolfsson2004,Ashmore2006,Zaninetti2015a,Ashmore2016}
and Table \ref{cosmologyvalues}.

A careful examination of equation (\ref{timegrb}) 
which gives the time duration for a GRB 
isolates five  fixed basic parameters 
which are  $\alpha$, $d$, $R_0$, $V_0$,$t_0$ and
two random parameters are , $L$,$D_L$.
In order to allow  the simulation 
the  five  fixed parameters are reported 
in Table \ref{tablesimple}.
\begin{table}[ht!]
\caption
{
The fixed  parameters    
of the simple model. 
}
 \label{tablesimple}
\begin{center}
\[
\begin{array}{cc}
\hline
Parameter       & value       \\ 
 \hline
t_0~ (s)             &  0.1     \\
R_0~ (pc)            &  0.06    \\
V_0~  (\frac{km}{s}) &  200000  \\
\alpha               &  0.8288  \\
d                      &  3.075   \\
\hline 
\end{array}
\]
\end{center}
\end{table}

The    variable  $L$ is  generated 
in a random way  which  follows a TL PDF
with the following meanings:
$L_l$ lower luminosity, 
$L_u$ upper  luminosity 
and 
$L^*$ the scale, see Table \ref{parameterslum}.
\begin{table}[ht!]
\caption
{
The   parameters  of the simulation 
for  $L$  in $\Lambda$CDM and 
plasma cosmology.
}
\label{parameterslum}
\begin{center}
\[
\begin{array}{cccc}
\hline
Parameter  &   \Lambda\!CDM~cosmology & Plasma~cosmology \\
\hline
 x_l=\frac{L_l}{10^{51} erg\,s^{-1}}   &  8.11 \,10^{-6}  &  1.77\,10^{-9} \\
 x_u=\frac{L_u}{10^{51} erg\,s^{-1}}   &  4.05 \,10^{-2}  &  2.36\,10^{-3} \\
m  = \frac{L^*}{10^{51} erg\,s^{-1}}   &  4.05 \,10^{-5}  &  5.9\,10^{-9}   \\
 \sigma  & 2.0   &   1.8 \\
\hline
\end{array}
\]
\end{center}
\end{table}

The theoretical luminosity at $t=t_0$, $L_0$,
is obtained inserting in equation 
(\ref{lsimpleastro}) the number density $n_0$ 
for which $L=L_0$. 
  
The    variable  $D_L$, 
in the case of $\Lambda$CDM or plasma cosmology, 
 is  generated 
in a random way  which  follows a TL PDF
with the following meanings:
$D_{L, l} $ lower  luminosity distance, 
$L_{L,u}  $ upper  luminosity distance 
and 
$m$  the scale , see Table \ref{parametersdl}.

\begin{table}[ht!]
\caption
{
The   parameters  of the simulation 
for  $D_L$  in $\Lambda$CDM and 
plasma cosmology.
}
\label{parametersdl}
\begin{center}
\[
\begin{array}{cccc}
\hline
Parameter  &   \Lambda\!CDM~cosmology & Plasma~cosmology \\
\hline
 D_{L,l}  \, (Mpc)       & 3                       &  1        \\
 D_{L,u}  \, (Mpc)       & 40000                   &  5000      \\
m         \, (Mpc)       &  700                    &  181.45    \\
 \sigma                  &  1.8                    &  1.19     \\
\hline
\end{array}
\]
\end{center}
\end{table}

\begin{figure}
\begin{center}
\includegraphics[width=10cm]{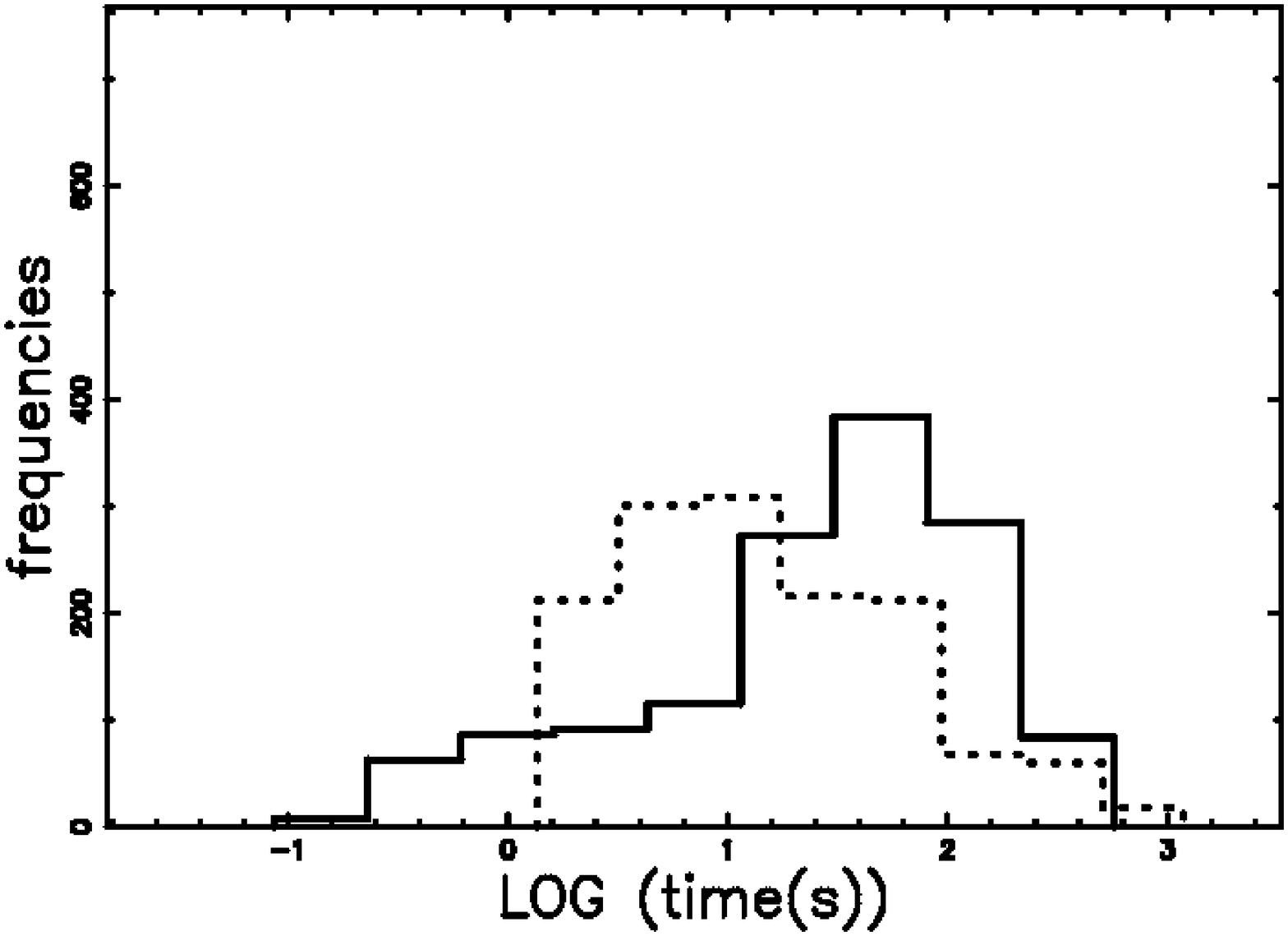}
\end{center}
\caption
{
Frequencies distribution (linear scale)) of $T_{90}$  
(decimal logarithm scale) 
for GRBs (full line) 
and theoretical simulation in the case of 
the $\Lambda$CDM cosmology (dotted line) 
with parameters as in Tables \ref{parameterslum}
and \ref{parametersdl}.
}
\label{histodue90_standard}
\end{figure}

\begin{figure}
\begin{center}
\includegraphics[width=10cm]{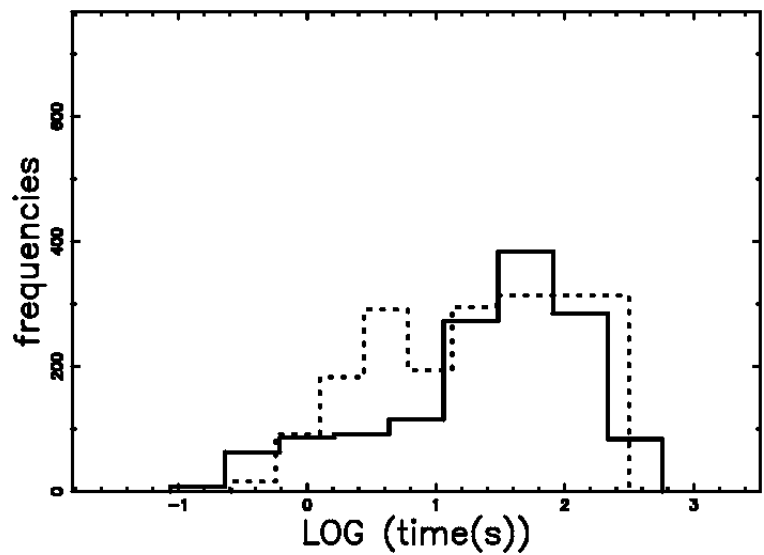}
\end{center}
\caption
{
Frequencies distribution 
(linear scale) of $T_{90}$  
(decimal logarithm scale) 
for GRBs (full line) 
and theoretical simulation in the case of 
the plasma  cosmology (dotted line)
with parameters as in Tables \ref{parameterslum}
and \ref{parametersdl}.
}
\label{histodue90_plasma}
\end{figure}

Once a sequence of theoretical times of duration is obtained 
according to equation (\ref{timegrb}),
see results in Table \ref{table_time_simple},  
we compare  the observed and simulated data,  
see Figures 
\ref{histodue90_standard} 
and 
\ref{histodue90_plasma}.

\begin{table}
 \caption[]
{
Theoretical time duration     
for GRBs in seconds.
}
 \label{table_time_simple}
 \[
 \begin{array}{lll}
 \hline
Parameter       &  \Lambda\!CDM~cosmology  & Plasma~cosmology\\ \noalign{\smallskip}
 \hline
 \noalign{\smallskip}
elements                & 1405         & 1405 \\
maximum~  (s)           & 3875.1       & 433     \\
mean~     (s)           & 36.2         & 16.84     \\
minimum~  (s)           & 0.828        & 0.162     \\
standard ~deviation (s) & 140.2        & 48.22     \\  
 \hline
 \end{array}
 \]
 \end {table}

The theoretical reason which allows
to fit the ratio  of two lognormal PDFs with another 
lognormal  is outlined in Section \ref{ratiolognormals}.

\section{Relativistic case}

\label{secrelativistic} 
The density, $\rho$,
of the ISM at a distance $r$ from the
SN is here  modeled  by  a
Plummer-like   profile, see \cite{Plummer1911},
\begin{equation}
\rho(r;R_{flat}) =
\rho_c  ({\frac {R_{flat}}{({R_{flat}^2 +r^2})^{1/2}}}  )^{\eta}
\label{densitaplummerflat}
\nonumber
\quad .
\end{equation}
where $r$  is the distance from the center,
$\rho$     is the density,
$\rho_c$   is the density at the center,
$R_{flat}$ is the distance before  which the density is nearly constant,
and $\eta$ is the power law exponent at large values of $r$.
The following transformation, $R_{flat}= \sqrt{3} b $,
gives 
\begin{equation}
\rho(r;b) =
\rho_c \bigl ( \frac{1}{1+ \frac{1}{3}\,{\frac {{r}^{2}}{{b}^{2}}}}
 \bigr ) ^{\eta/2}
\quad .
\label{densitaplummer}
\quad
\end{equation}
The  total  mass $M(r;b)$
comprised between
0 and  $r$,  when  $\eta=6$, 
is
\begin{eqnarray}
M(r;b) = \int_0^r   4 \pi r^2   \rho(r;b) dr
= 
\frac{3}{2}\,{\frac {\rho\_c,\pi\,{b}^{3}}{ ( 3\,{b}^{2}+{r}^{2}    ) ^{
2}}  \Bigl  ( 9\,\arctan    ( \frac{1}{3}\,{\frac {r\sqrt {3}}{b}}    ) 
\sqrt {3}{b}^{4}
}
\nonumber \\
{+6\,\arctan    ( \frac{1}{3}\,{\frac {r\sqrt {3}}{b}}
    ) \sqrt {3}{b}^{2}{r}^{2}}
{+\arctan    ( \frac{1}{3}\,{\frac {r\sqrt {3
}}{b}}    ) \sqrt {3}{r}^{4}-9\,{b}^{3}r+3\,b{r}^{3} \Bigr   ) }
\quad .
\end{eqnarray}
The relativistic  conservation of momentum,
see \cite{French1968,Zhang1997,Guery2010},
states that 
\begin{equation}
M(R_0;b) \gamma_0 \beta_0 = M(r;b) \gamma \beta
\quad ,
\nonumber
\end{equation}
with 
\begin{equation}
\gamma_0 = \frac{1} {
\sqrt{1-\beta_0^2}
}
\quad ; \qquad
\gamma = \frac{1} {
\sqrt{1-\beta^2}
}
\quad ,
\nonumber
\end {equation}
and
\begin{equation}
\beta_0 =\frac{V_0}{c}
\quad ; \qquad
\beta =\frac{v}{c} 
\quad , \nonumber
\end{equation}
where 
$R_0$  is the initial radius of the advancing sphere,
$V_0$  is the initial velocity at $R_0$  and
$c$    is the light  velocity.

The relativistic  conservation of momentum
for a Plummer profile with $\eta=6$ is 
\begin{equation}
\frac{AN}{AD} = \frac{BN}{BD}
\quad , 
\label{eqndiffrelplummer}
\end{equation}
\begin{eqnarray}
AN=
3\,  \,\pi\,{b}^{3}  \Bigl  ( 9\,\arctan   ( \frac{1}{3}\,{\frac {r   ( t
    ) \sqrt {3}}{b}}    ) \sqrt {3}{b}^{4}+6\,\arctan   ( \frac{1}{3}
\,{\frac {r   ( t    ) \sqrt {3}}{b}}    ) \sqrt {3}{b}^{2}
   ( r   ( t    )     ) ^{2}
\nonumber  \\
+\arctan   ( \frac{1}{3}\,{\frac {r
   ( t    ) \sqrt {3}}{b}}    ) \sqrt {3}   ( r   ( t
    )     ) ^{4}-9\,{b}^{3}r   ( t    ) +3\,b   ( r
   ( t    )     ) ^{3}   \Bigr ) {\frac {\rm d}{{\rm d}t}}r
   ( t    ) 
\end {eqnarray}
\begin{equation}
AD=
2\, \left( 3\,{b}^{2}+ \left( r \left( t \right)  \right) ^{2}
 \right) ^{2}c\sqrt {1-{\frac { \left( {\frac {\rm d}{{\rm d}t}}r
 \left( t \right)  \right) ^{2}}{{c}^{2}}}}
\quad  , 
\end{equation}

\begin{eqnarray}
BN=
3\,  \,\pi\,{b}^{3}  \Bigl  ( 9\,\arctan    ( \frac{1}{3}\,{\frac {R_{{0}}
\sqrt {3}}{b}}    ) \sqrt {3}{b}^{4}+6\,\arctan    ( \frac{1}{3}\,{\frac 
{R_{{0}}\sqrt {3}}{b}}    ) \sqrt {3}{b}^{2}{R_{{0}}}^{2}
\nonumber  \\
+\arctan
    ( \frac{1}{3}\,{\frac {R_{{0}}\sqrt {3}}{b}}    ) \sqrt {3}{R_{{0}}}^
{4}
-9\,{b}^{3}R_{{0}}+3\,b{R_{{0}}}^{3}  \Bigr  ) \beta_{{0}}
\quad  , 
\end{eqnarray}
\begin{equation}
BD=
2\, \left( 3\,{b}^{2}+{R_{{0}}}^{2} \right) ^{2}\sqrt {1-{\beta_{{0}}}^
{2}}
\quad .
\end{equation}

The relativistic transfer  of  energy
through a surface, $A$, is
\begin{equation}
L_{m,r}  = A \gamma^2 ( \rho c^2 +p ) v
\quad ,
\end{equation}
where $p$  is the pressure here assumed to be  $p$=0; 
see Eq. A31  in \cite{deyoung} or Eq. (43.44) in
\cite{Mihalas2013}.

The astrophysical version 
of the relativistic transfer 
of energy ( the luminosity) at $t=t_0$  is
\begin{equation}
L=  5.37\,{\frac {{\it n_0}\,{{\it R_{0,pc}}}^{2}\beta_0}{1-{\beta_0}^{
2}}}
 10^{32}
\, W   \quad ,
\label{lrelativisticastro}
\end{equation}
where 
$R_{0,pc}$ is the radius    at $t=t_0$ in pc.

In the case of a spherical cold expansion
\begin{equation}
L_{m,r} = 4 \pi r(t)^2 \frac{1}{1-\beta(t)^2} \rho(t) c^3 \beta(t)
\quad .
\end{equation}
We now assume the  following  power law behavior  
for the
density in the advancing thin layer
\begin{equation}
\rho(t) = \rho_0 (\frac{t_0} {t})^d
\quad ,
\end{equation}
and we obtain
\begin{equation}
L_{m,r} =
4 \pi r(t)^2
\frac{1}{1-\beta(t)^2}
\rho_0 (\frac{t_0} {t})^d
c^3 \beta(t)
\quad .
\label{luminosityrel}
\end{equation}
We can now  derive $L_{m,r}$ in two ways: 
(i)  from a numerical evaluation of r(t) and  v(t),
(ii) from a Taylor series  of
$L_{m,r}(t)$ of the type
\begin{equation}
L_{m,r}(t) =
\sum _{n=0}^{3}a_{{n}}{(t-t_0)}^{n}
\quad .
\label{luminosityseries}
\end{equation}
The first two coefficients are
\begin{eqnarray}
a_0=&  
-\rho_0\,4\,\pi\,{R_{{0}}}^{2} \left( 
{\frac {t_{{0}}}{t}} \right) ^{d}{c}^{3}
\beta_{{0}}
  \nonumber \\
a_1 =&  
\rho_0\,\frac{A1N}{A1D} 
  \nonumber    
\quad ,
\end{eqnarray}
where 
\begin{eqnarray}
A1N =
-8\,\pi\,R_{{0}} \left( {\frac {t_{{0}}}{t}} \right) ^{d}{c}^{4}{\beta
_{{0}}}^{2} \Biggl ( 27\,\sqrt {3}{b}^{6}\arctan \left( \frac{1}{3}\,{\frac {r_{
{0}}\sqrt {3}}{b}} \right) 
\nonumber  \\
+27\,\sqrt {3}{b}^{4}{R_{{0}}}^{2}\arctan
 \left( \frac{1}{3}\,{\frac {R_{{0}}\sqrt {3}}{b}} \right) 
+9\,\sqrt {3}{b}^{2
}{R_{{0}}}^{4}\arctan \left( \frac{1}{3}\,{\frac {R_{{0}}\sqrt {3}}{b}}
 \right) 
\nonumber  \\
+\sqrt {3}{R_{{0}}}^{6}\arctan \left( \frac{1}{3}\,{\frac {R_{{0}}
\sqrt {3}}{b}} \right) -36\,{R_{{0}}}^{3}{\beta_{{0}}}^{2}{b}^{3}-27\,
{b}^{5}R_{{0}}
\nonumber  \\
-36\,{R_{{0}}}^{3}{b}^{3}+3\,b{R_{{0}}}^{5} \Biggr) 
\quad ,
\end{eqnarray}
and
\begin{eqnarray}
A1D =
\Biggl  ( 27\,\sqrt {3}{b}^{6}\arctan   ( \frac{1}{3}\,{\frac {R_{{0}}\sqrt {
3}}{b}}   ) +27\,\sqrt {3}{b}^{4}{R_{{0}}}^{2}\arctan   ( \frac{1}{3}\,
{\frac {R_{{0}}\sqrt {3}}{b}}   ) 
\nonumber  \\
+9\,\sqrt {3}{b}^{2}{R_{{0}}}^{4
}\arctan   ( \frac{1}{3}\,{\frac {R_{{0}}\sqrt {3}}{b}}   ) +\sqrt {3}{
R_{{0}}}^{6}\arctan   ( \frac{1}{3}\,{\frac {R_{{0}}\sqrt {3}}{b}}   ) 
-27\,{b}^{5}R_{{0}}+3\,b{R_{{0}}}^{5} \Biggr   ) \times
\nonumber \\
    ( 1- {\beta_{{0}}}^{2
}   )
\quad  . 
\end{eqnarray}

\begin{figure*}
\begin{center}
\includegraphics[width=7cm]{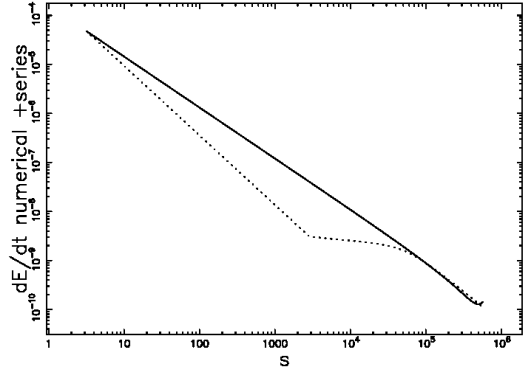}
\end {center}
\caption
{
Numerical $L_{m,r}$
computed according to  equation (\ref{luminosityrel}) (dotted line)
and series solution of order 7 as
given by equation (\ref{luminosityseries}) (full  line).
Data as in Table \ref{datafitrelplummer}.
}
\label{dedt_numerical_series}
    \end{figure*}

Figure \ref{dedt_numerical_series} compares
the numerical  solution for the luminosity and
the series expansion for  the
luminosity about the  ordinary point $t_0$.

\begin{table}
\caption
{
Numerical values of the parameters
used in the Plummer  relativistic solution.
}
 \label{datafitrelplummer}
 \[
 \begin{array}{c}
 \hline
 \hline
 \noalign{\smallskip}
 parameters      \\
  t_0=1 \times 10^{-7}~\mathrm{yr}~or~t_0=3.15~\mathrm{seconds}~\\
  R_0=0.01~\mathrm{pc} \\
 \beta_0=0.666 \\
  b=0.028~ \mathrm{pc}\\
  d=1 \\
\noalign{\smallskip} \hline
 \end{array}
 \]
 \end {table}
The relativistic theory is now applied to
\secondogrb
at 0.3-10 kev  in the time interval
$10^{-5}-3$ days,
see \cite{Jakobsson2006},
with data available at
\url{http://www.swift.ac.uk/xrt_curves/150314}.

\begin{figure}
\begin{center}
\includegraphics[width=6cm]{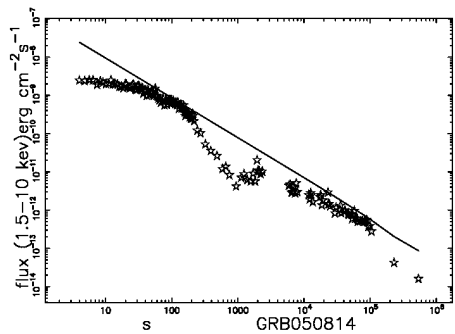}
\end {center}
\caption
{
The XRT flux  of \secondogrb
at 0.3-10 kev  (empty stars)  and
theoretical curve as given by the 
relativistic numerical model,
see equation  (\ref{luminosityrel})
(full line) , no absorption.
}
\label{flux_obs_theo}
    \end{figure}

The  theoretical flux  without 
absorption is given by 
equation   \ref{fluxobserved} 
and Figure \ref{flux_obs_theo} 
reports the comparison of the 
theoretical flux and  the observed one.

\begin{figure}
\begin{center}
\includegraphics[width=6cm]{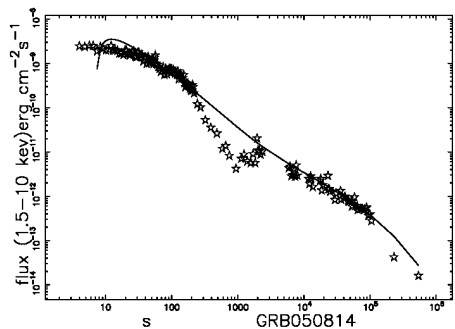}
\end {center}
\caption
{
The XRT flux  of \secondogrb
at 0.3-10 kev  (empty stars)  and
theoretical curve as given by the relativistic   numerical model,
see equation  (\ref{luminosityrel}),
in presence of absorption as given by
equation  (\ref{fluxobservedtau})
(full line).
}
\label{grb050814_rel_plummer}
    \end{figure}

The presence of the 
absorption can be modeled 
as
\begin{equation}
F = \frac{L}{4 \pi D^2}
(1-e^{-\tau_{\nu}(t)})
\quad ,
\label{fluxobservedtau}
\end{equation}
where $\tau_{nu}(t)$   is the optical thickness
here assumed to be dependent from the time.
As a model for   ${\tau_{\nu}}$ as function of time
we select a logarithmic polynomial approximation,
of degree 5 , see  
\cite{Zaninetti2015c} for more details.
Figure \ref{grb050814_rel_plummer} 
reports the LC for  the relativistic case 
with absorption  for \secondogrb.

\subsection{The relativistic simulation for time duration}

In the relativistic case the theoretical  luminosity
is  provided by a 
series solution of order 7, see equation (\ref{luminosityseries}),
The fixed parameters adopted for the simulation  of
duration time  are  reported in Table  \ref{timerelplummer}.

\begin{table}
\caption
{
Numerical values of the parameters
used in the Plummer  relativistic 
simulation for duration
time of GRBs.
}
 \label{timerelplummer}
 \[
 \begin{array}{c}
 \hline
 \hline
 \noalign{\smallskip}
 parameters      \\
  t_0=0.1 \mathrm{s}\\
  R_0=0.06~\mathrm{pc} \\
 \beta_0=\frac{2}{3} \\
 b=0.028~ \mathrm{pc} \\
  d=1.04 \\
\noalign{\smallskip} \hline
 \end{array}
 \]
 \end {table}

The    random luminosity  $L$ is  generated 
according  to a  TL PDF,
see  Table \ref{parameterslumrel}.

\begin{figure}
\begin{center}
\includegraphics[width=10cm]{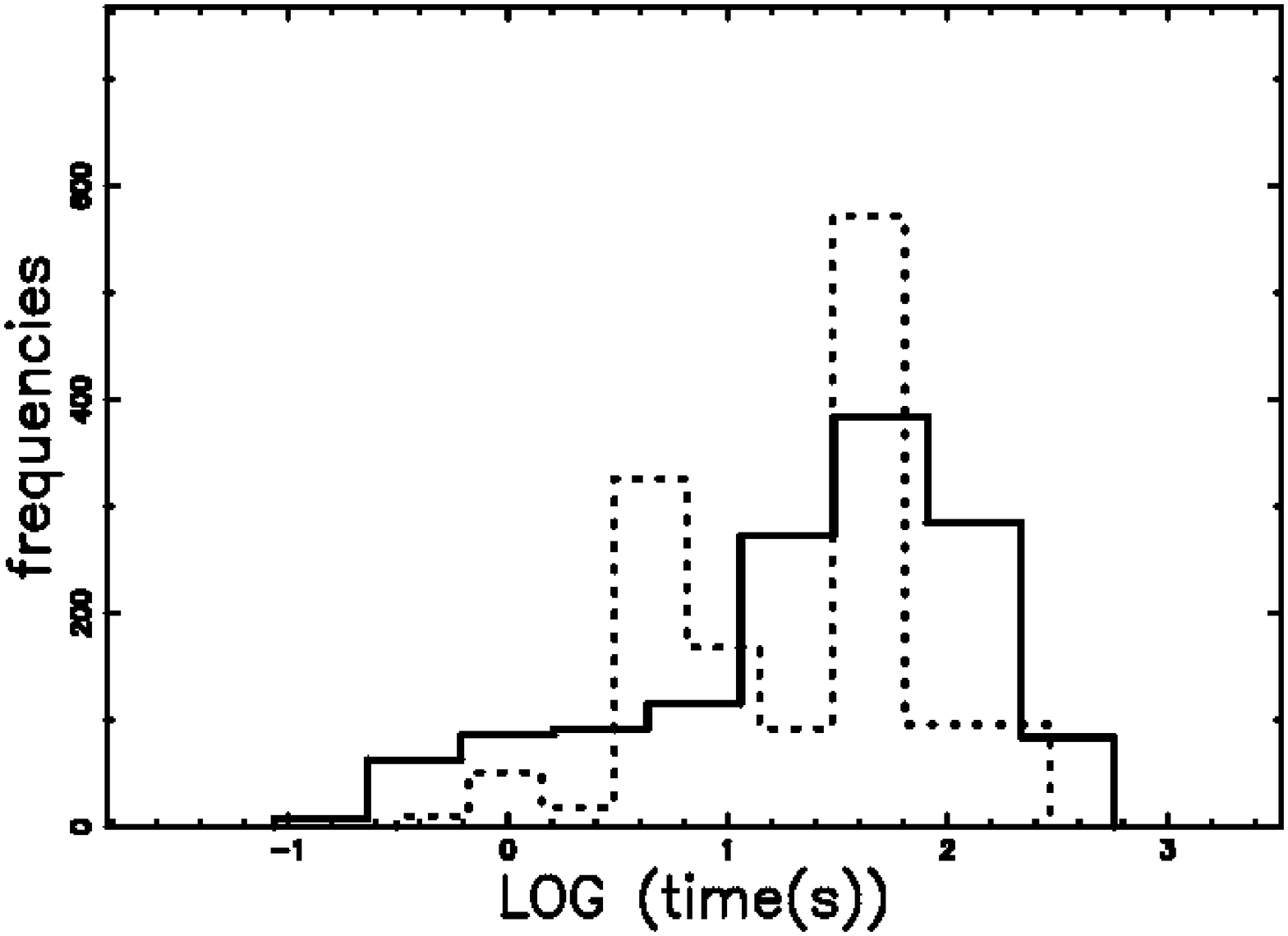}
\end{center}
\caption
{
Frequencies distribution (linear scale)) of 
$T_{90}$  
(decimal logarithm scale) 
for GRBs (full line) 
and theoretical simulation in the case of 
the $\Lambda$CDM cosmology (dotted line)
with parameters as in Tables \ref{parameterslum}
and \ref{parametersdl}.
The   model is relativistic, 
see equation  (\ref{luminosityrel})
(full line) without  absorption.
}
\label{histodue90_standard_rel}
\end{figure}

\begin{figure}
\begin{center}
\includegraphics[width=10cm]{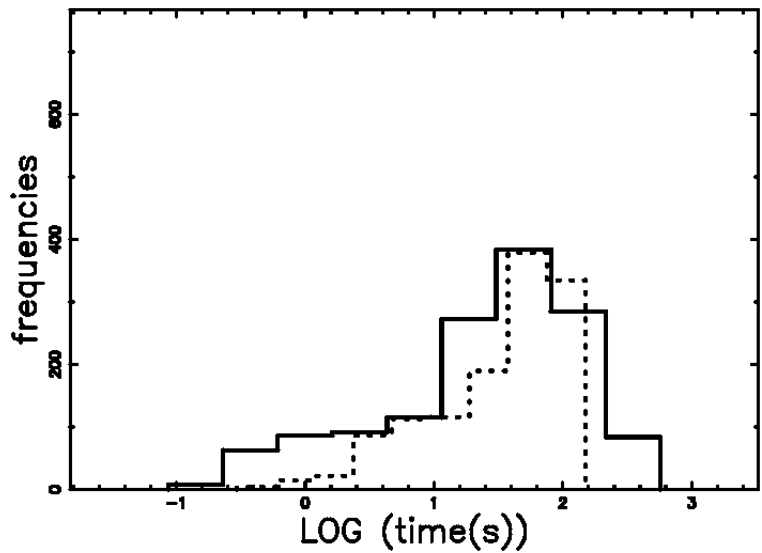}
\end{center}
\caption
{
Frequencies distribution (linear scale)) of $T_{90}$  
(decimal logarithm scale) 
for GRBs (full line) 
and theoretical simulation in the case of 
the plasma  cosmology (dotted line)
in a relativistic framework 
with parameters as in Tables \ref{parameterslumrel}
and \ref{parametersdlrel}.
}
\label{histodue90_plasma_rel}
\end{figure}

\begin{table}[ht!]
\caption
{
The   parameters  of the relativistic simulation 
for  $L$  in $\Lambda$CDM and 
plasma cosmology.
}
\label{parameterslumrel}
\begin{center}
\[
\begin{array}{cccc}
\hline
Parameter  &   \Lambda\!CDM~cosmology & Plasma~cosmology \\
\hline
 x_l=\frac{L_l}{10^{51} erg\,s^{-1}}   &  4.11 \,10^{-8}  &  5.9\,10^{-12}                     \\
 x_u=\frac{L_u}{10^{51} erg\,s^{-1}}   &  4.05 \,10^{-2}  &  2.95\,10^{-5}                   \\
m  = \frac{L^*}{10^{51} erg\,s^{-1}}   &  9.8  \,10^{-4}  &  5.9\,10^{-9}   \\
 \sigma  & 1.42   &   1.42 \\
\hline
\end{array}
\]
\end{center}
\end{table}
The value of $n_0$ which allows to generate 
$L_0$ in a random way is deduced by $L=L_0$ 
with 
$L$ as evaluated in equation (\ref{lrelativisticastro}).
The    distance luminosity, $D_L$, 
in the case of $\Lambda$CDM or plasma cosmology, 
i randomly generated according to  
a TL PDF, see Table \ref{parametersdlrel}.
\begin{table}[ht!]
\caption
{
The   parameters  of the relativistic simulation 
for  $D_L$  in $\Lambda$CDM and 
plasma cosmology.
}
\label{parametersdlrel}
\begin{center}
\[
\begin{array}{cccc}
\hline
Parameter  &   \Lambda\!CDM~cosmology & Plasma~cosmology \\
\hline
 D_{L,l}  \, (Mpc)       & 3                       &  1.19        \\
 D_{L,u}  \, (Mpc)       & 38000                   &  6225      \\
m         \, (Mpc)       &  1000                    &  181.45    \\
 \sigma                  &  1.4                   &  1.19     \\
\hline
\end{array}
\]
\end{center}
\end{table}
The  sequence of theoretical times of duration is obtained 
in a numerical way,
see results in Table \ref{table_time_relativistic}.
The  observed and simulated data  are displayed in   
Figures 
\ref{histodue90_standard_rel} 
and 
\ref{histodue90_plasma_rel}.

\begin{table}
 \caption[]
{
Theoretical time duration      
for GRBs in seconds in a relativistic framework.
}
 \label{table_time_relativistic}
 \[
 \begin{array}{lll}
 \hline
Parameter       &  \Lambda\!CDM~cosmology  & Plasma~cosmology\\ \noalign{\smallskip}
 \hline
 \noalign{\smallskip}
elements                & 1405         & 1405 \\
maximum~  (s)           & 400          & 200    \\
mean~     (s)           & 34.94        & 53.46     \\
minimum~  (s)           & 0.2          & 0.2     \\
standard ~deviation (s) & 77.32        & 56.88     \\  
 \hline
 \end{array}
 \]
 \end {table}

\section{Conclusions}

{\bf Extra-galactic versus Galactic origin}

The direction of arrival of GRBs   shows an isotropic 
universe  with a percentage error  
of $2.4\%$.
This means that the GRBs are originated in external galaxies
in a random way  as  suggested by  \cite{Balazs2010}.
 
{\bf Truncated lognormal distribution}

The lognormal PDF is usually adopted to model
the duration time of GRBs,  
see \cite{Horvath2016}.
The TL PDF improves the reliability of
the fit, see Table  \ref{table_time90}.
Careful attention should be paid to the fact
that a two-Gaussian and  a three-Gaussian fit
are  also used to model the duration times for GRBs,
see \cite{Tarnopolski2015}.

{\bf A simple model}

The theoretical time of duration  
can be derived from the light curve for GRBs,
A power law  approximation for the time 
of expansion of a shell,
see equation (\ref{rpower}), 
coupled with a power law behavior 
for the density as function of the radius, see 
equation (\ref{rhor}),
allows to derive a formula
for the theoretical  luminosity, 
see equation (\ref{eqnl}). 
A theoretical time of duration is obtained,
see equation  (\ref{timegrb}) which contains
an evaluation of the luminosity and 
the luminosity distance.
The cosmological evaluation of the luminosity distance
is given both in 
$\Lambda$CDM cosmology
and plasma cosmology.
The results are given in 
Table \ref {table_time_simple},
Figures 
\ref{histodue90_standard} 
and 
\ref{histodue90_plasma}.

{\bf Relativistic model} 

The temporal  evolution of a SN in a  medium
of the Plummer  type, $\eta=6$,   can be found by applying the
conservation of relativistic momentum in the thin layer
approximation.
This  relativistic invariant is  evaluated as
a differential equation of the first order, see
equation  (\ref{eqndiffrelplummer}).
Two different relativistic solutions   
for the theoretical luminosity  as a function
of time  are derived:
(i) a numerical  solution,
see Eq.~(\ref{luminosityrel});
(ii) a series solution, see Eq.~(\ref{luminosityseries}),
which has a limited
range of validity,
$10^{-1}\mathrm{ s} < \, t < 10^{6} \, \mathrm{s} $.

The coupling of the previous series
with a logarithmic  polynomial approximation
allows to model fine details such as the
oscillation in LC visible at $\approx$ 1000 s in
\grb, see  Figure \ref{grb050814_rel_plummer}.

The relativistic time of duration is reported in 
in Table \ref{table_time_relativistic},
Figures 
\ref{histodue90_standard_rel} 
and 
\ref{histodue90_plasma_rel}.

An enlightening example of such relativistic simulation
is absence  of bursts  at $t\approx 15.8\,s$ 
visible in Figure \ref{histodue90_standard_rel}.
This means that the claimed boundary 
at $t \approx 2\,s$ which divides 
the short   by  the long bursts, see \cite{Berger2014}, 
is due to random events rather than 
two different kind of bursts.  

\section*{Acknowledgments}

This work made use of data supplied by the UK 
Swift Science Data Centre at
the University of Leicester.


\begin{thebibliography}{10}
\expandafter\ifx\csname url\endcsname\relax
  \def\url#1{{\tt #1}}\fi
\expandafter\ifx\csname urlprefix\endcsname\relax\def\urlprefix{URL }\fi
\providecommand{\eprint}[2][]{\url{#2}}

\bibitem{Dado2016}
{Dado} S and {Dar} A 2016 {Critical test of gamma-ray burst theories} {\em
  \prd\/} {\bf 94}(6) 063007 (\textit{Preprint} \eprint{1603.06537})

\bibitem{Willingale2017}
{Willingale} R and {M{\'e}sz{\'a}ros} P 2017 {Gamma-Ray Bursts and Fast
  Transients - Multi-wavelength Observations and Multi-messenger Signals} {\em
  \ssr\/} pp 1--24

\bibitem{Asano2016}
{Asano} K and {M{\'e}sz{\'a}ros} P 2016 {Ultrahigh-energy cosmic ray production
  by turbulence in gamma-ray burst jets and cosmogenic neutrinos} {\em \prd\/}
  {\bf 94}(2) 023005 (\textit{Preprint} \eprint{1607.00732})

\bibitem{Moharana2016}
{Moharana} R, {Razzaque} S, {Gupta} N and {M{\'e}sz{\'a}ros} P 2016
  {High-energy neutrinos from the gravitational wave event GW150914 possibly
  associated with a short gamma-ray burst} {\em \prd\/} {\bf 93}(12) 123011
  (\textit{Preprint} \eprint{1602.08436})

\bibitem{Netchitailo_2017}
{Netchitailo} V~S 2017 Burst astrophysics {\em Journal of High Energy Physics,
  Gravitation and Cosmology\/} {\bf 03}(02), 157

\bibitem{Bhat2016}
{Narayana Bhat} P, {Meegan} C~A, {von Kienlin} A and {Paciesas} W~S 2016 {The
  Third Fermi GBM Gamma-Ray Burst Catalog: The First Six Years} {\em \apjs\/}
  {\bf 223} 28 (\textit{Preprint} \eprint{1603.07612})

\bibitem{Berger2014}
{Berger} E 2014 {Short-Duration Gamma-Ray Bursts} {\em \araa\/} {\bf 52}, 43
  (\textit{Preprint} \eprint{1311.2603})

\bibitem{Tarnopolski2015}
{Tarnopolski} M 2015 {Analysis of Fermi gamma-ray burst duration distribution}
  {\em \aap\/} {\bf 581} A29 (\textit{Preprint} \eprint{1506.07324})

\bibitem{Horvath2016}
{Horv{\'a}th} I and {T{\'o}th} B~G 2016 {The duration distribution of Swift
  Gamma-Ray Bursts} {\em \apss\/} {\bf 361} 155 (\textit{Preprint}
  \eprint{1604.00887})

\bibitem{Zaninetti2016c}
{Zaninetti} L 2016 {The Truncated Lognormal Distribution as a Luminosity
  Function for SWIFT-BAT Gamma-Ray Bursts} {\em Galaxies\/} {\bf 4}, 57
  (\textit{Preprint} \eprint{1611.01650})

\bibitem{Kolmogoroff1941}
{Kolmogoroff} A 1941 Confidence limits for an unknown distribution function
  {\em The Annals of Mathematical Statistics\/} {\bf 12}(4), 461 ISSN 00034851

\bibitem{Smirnov1948}
{Smirnov} N 1948 Table for estimating the goodness of fit of empirical
  distributions {\em The Annals of Mathematical Statistics\/} {\bf 19}(2), 279
  ISSN 00034851

\bibitem{Massey1951}
{Massey} Frank~J J 1951 The kolmogorov-smirnov test for goodness of fit {\em
  Journal of the American Statistical Association\/} {\bf 46}(253), 68

\bibitem{Zaninetti2011a}
{Zaninetti} L 2011 {Time-dependent models for a decade of SN 1993J} {\em
  \apss\/} {\bf 333}, 99

\bibitem{deyoung}
{De Young} D~S 2002 {\em {The physics of extragalactic radio sources}\/}
  (Chicago: University of Chicago Press)

\bibitem{Evans2007}
{Evans} P~A, {Beardmore} A~P and {Page} K~L 2007 {An online repository of
  Swift/XRT light curves of {$\gamma$}-ray bursts} {\em \aap\/} {\bf 469}, 379
  (\textit{Preprint} \eprint{0704.0128})

\bibitem{Evans2009}
{Evans} P~A, {Beardmore} A~P and {Page} K~L 2009 {Methods and results of an
  automatic analysis of a complete sample of Swift-XRT observations of GRBs}
  {\em \mnras\/} {\bf 397}, 1177 (\textit{Preprint} \eprint{0812.3662})

\bibitem{Krimm2016}
{Krimm} H~A, {Barthelmy} S~D, {Cummings} J~R, {D'Avanzo} P, {Gehrels} N, {Lien}
  A~Y, {Markwardt} C~B, {Palmer} D~M, {Sakamoto} T, {Stamatikos} M and
  {Ukwatta} T~N 2016 {GRB 161214B: Swift-BAT refined analysis.} {\em GRB
  Coordinates Network\/} {\bf 20270}

\bibitem{Zaninetti2016a}
{Zaninetti} L 2016 Pade approximant and minimax rational approximation in
  standard cosmology {\em Galaxies\/} {\bf 4}(1), 4 ISSN 2075-4434
  \urlprefix\url{http://www.mdpi.com/2075-4434/4/1/4}

\bibitem{Brynjolfsson2004}
{Brynjolfsson} A 2004 {Redshift of photons penetrating a hot plasma} {\em
  arXiv:astro-ph/0401420\/}

\bibitem{Ashmore2006}
{Ashmore} L 2006 Recoil between photons and electrons leading to the hubble
  constant and cmb {\em Galilean Electrodynamics\/} {\bf 17}(3), 53

\bibitem{Zaninetti2015a}
{Zaninetti} L 2015 {On the Number of Galaxies at High Redshift} {\em
  Galaxies\/} {\bf 3}, 129

\bibitem{Ashmore2016}
{Ashmore} L 2016 A relationship between dispersion measure and redshift derived
  in terms of new tired light. {\em Journal of High Energy Physics, Gravitation
  and Cosmology\/} {\bf 2}, 512

\bibitem{Plummer1911}
{Plummer} H~C 1911 {On the problem of distribution in globular star clusters}
  {\em \mnras\/} {\bf 71}, 460

\bibitem{French1968}
{{French}, AP} 1968 {\em {Special Relativity}\/} (New~York: {CRC})

\bibitem{Zhang1997}
{Zhang} Y 1997 {\em Special Relativity and Its Experimental Foundations\/}
  (Singapore: World Scientific) ISBN 9789810227494

\bibitem{Guery2010}
Gu{\'e}ry-Odelin D and Lahaye T 2010 {\em Classical Mechanics Illustrated by
  Modern Physics: 42 Problems with Solutions\/} (London: Imperial College
  Press) ISBN 9781848164802

\bibitem{Mihalas2013}
Mihalas D and Mihalas B 2013 {\em Foundations of Radiation Hydrodynamics\/}
  Dover Books on Physics (New York: Dover Publications) ISBN 9780486135885
  \urlprefix\url{http://books.google.it/books?id=GVK8AQAAQBAJ}

\bibitem{Jakobsson2006}
{Jakobsson} P, {Levan} A and {Fynbo} J~P 2006 {A mean redshift of 2.8 for Swift
  gamma-ray bursts} {\em \aap\/} {\bf 447}, 897 (\textit{Preprint}
  \eprint{astro-ph/0509888})

\bibitem{Zaninetti2015c}
{Zaninetti} L 2015 {Relativistic Scaling Laws for the Light Curve in
  Supernovae} {\em Applied Physics Research\/} {\bf 7}, 48

\bibitem{Balazs2010}
{Bal{\'a}zs} L~G, {Vavrek} R, {M{\'e}sz{\'a}ros} A, {Horv{\'a}th} I, {Bagoly}
  Z, {Veres} P and {Tusn{\'a}dy} G 2010 {Is sky distribution of gamma-ray
  bursts random?} {\em Astrophysical Bulletin\/} {\bf 65}, 277

\end{thebibliography}
\providecommand{\newblock}{}

\end{document}